\documentclass[12pt]{article}
\usepackage{amssymb}

\newcommand{\beq}[1]{\begin{equation}\label{#1}}
\newcommand{\eeq}{\end{equation}}
\newcommand{\bear}[1]{\begin{eqnarray}\label{#1}}
\newcommand{\ear}{\end{eqnarray}}
\newcommand{\nn}{\nonumber}

\renewcommand{\theequation}{\arabic{section}.\arabic{equation}}
\catcode`\@=11 \@addtoreset{equation}{section}\catcode`\@=12

 \def\barr{\left(\begin{array}}
 \def\earr{\end{array}\right)}

\newcommand{\N}{ {\mathbb N} }
\newcommand{\R}{ {\mathbb R} }

\newcommand{\diag}{ \mbox{\rm diag} }
\newcommand{\sign}{ \mbox{\rm sign} }
\newcommand{\e}{ \mbox{\rm e} }
\newcommand{\eps}{ \varepsilon }

\newcommand{\p}{\partial}

\newcommand{\tri}{\Delta}

\newcommand{\fnm}{\footnotemark}
\newcommand{\fnt}{\footnotetext}

\begin{document}

 \begin{center}
 \large \bf

Quantum billiards with branes  on product of Einstein spaces

 \end{center}

 \vspace{0.3truecm}

 \begin{center}

 \normalsize\bf

  V. D. Ivashchuk\fnm[1]\fnt[1]{e-mail: ivashchuk@mail.ru}

  \vspace{0.5truecm}

 \it Center for Gravitation and Fundamental
     Metrology, VNIIMS, Ozyornaya St., 46, Moscow 119361, Russia and

 \it Institute of Gravitation and Cosmology, Peoples' Friendship
     University of Russia,  Miklukho-Maklaya St.,6, Moscow 117198,
     Russia

 \end{center}

 \begin{abstract}

We consider a gravitational  model in dimension $D$ with several forms, $l$ scalar fields
and a $\Lambda$-term. We study cosmological-type  block-diagonal metrics defined on a product of an $1$-dimensional
interval and $n$ oriented Einstein spaces. As an
electromagnetic composite brane ansatz is adopted and certain
restrictions on the branes are imposed the conformally covariant
Wheeler-DeWitt (WDW) equation for the model is studied. Under
certain restrictions, asymptotic solutions to the WDW equation are
found in the limit of the formation of the billiard walls. These  solutions
reduce the problem to the so-called quantum billiard in  $(n +
l -1)$-dimensional hyperbolic space. Several examples of quantum billiards
in the model with  electric and magnetic  branes, e.g.
corresponding to hyperbolic Kac-Moody algebras, are considered.
In the case $n=2$ we find a set of basis asymptotic solutions
to the WDW equation and derive asymptotic solutions for the metric in the classical case.

\end{abstract}


 \large

  \section{Introduction}

In this paper we deal with  the  quantum billiard approach to
$D$-dimensional \\
cosmological-type  models defined on a
(warped) product manifold  $(u_{-},u_{+}) \times M_{1} \times \cdots \times M_{n}$, where
 $M_i$ is a smooth oriented Einstein manifold, $i=1,\ldots,n$.

 The billiard approach in classical gravity originally appeared in the
 dissertation of  Chitr\'e \cite{Chit} for an explanation of the
 BKL oscillations  \cite{BLK,BLK-1} while approaching to a spacelike singularity
 in the   Bianchi-IX model  \cite{Mis0}.
 In this description  a simple triangle billiard
 in the hyperbolic (Lobachevsky) space $H^2$ was used.
 The BKL-like  behavior near  a timelike singularity was studied in  \cite{Par1,Par2}.

 In \cite{GS},  the billiard approach for $D=4$ was extended to the quantum
 case (see also \cite{Kir-94}), i.e. the solutions to  the Wheeler-DeWitt
 (WDW) equation \cite{DeWitt} were  reduced
to the problem of finding the spectrum of the Laplace-Beltrami
operator on Chitr\'e's triangle billiard.

 Chitr\'e's billiard approach was generalized to a   multidimensional
 cosmological model with multicomponent anisotropic (``perfect'') fluid  \cite{IKM1,IKM1-1,IMb0}
defined on the product of $n$ Einstein factor spaces and 1-dimensional manifold.
 The search for an oscillating behavior near the singularity was
 reduced to the problem of proving the finiteness of  the billiard volume.
  At that time the quantum  billiard  approach to multidimensional cosmology was
 suggested in  \cite{IMb0,Kir-95,Kir-95-1}.

 The classical  billiard approach to  multidimensional models
 with  fields of forms and scalar fields in the presence of the $\Lambda$-term
 was suggested in  \cite{IMb1} along  lines suggested earlier in \cite{IMb0}.
 In ref. \cite{IMb1}  rather a general composite electromagnetic
ansatz for the fields of forms on a warped product of several Einstein manifolds and an 1-dimensional
base manifold $M_0$  was developed.
Reference   \cite{IMb1} contained  rather a general construction of the billiard approach
 for the description of the behavior of scale factors and scalar fields
near  either a spacelike or a timelike singularity, i.e. the metric in \cite{IMb1}
   $ds^2 = w du^2 + ...$, contained an arbitrary sign $w = \pm 1$ and  a coordinate $u$.
 Thus,  the paper  \cite{IMb1} was dealing with
cosmological-type solutions, e.g. cosmological,
spherically symmetric, and cylindrically symmetric ones. The metric
had a block-diagonal form. In  \cite{IMb1}
 the necessary condition for the formation of walls was
formulated in terms of inequalities for scalar products of the brane
vectors $U^s$:  $(U^s, U^s) > 0$
 and for the  so-called brane sign parameters: $\eps_s > 0$.
Inequalities on Kasner parameters, where formulated in terms
of linear functions $U^s(\alpha)$, which give either Kasner
 or oscillatory asymptotic  regimes near the singularity.
 Another advantages of the approach of \cite{IMb1}  was
 in dealing with a wide variety of signatures of Einstein factor space metrics (though restricted by
$\eps_s > 0$). It was shown that the curvatures of the
Einstein factor spaces and the $\Lambda$-term are irrelevant near
the singularity.

 Meanwhile the approach of ref. \cite{IMb1} had some restrictive points, since it was
dealing with block-diagonal metrics and   putting restrictions
on brane intersections (for branes corresponding to the same form
field) which guaranteed block-diagonal structure of
the stress-energy tensor. For some extension of these restrictions see ref. \cite{IMS}.

Some problems of the approach of ref. \cite{IMb1} were overcome in the
papers of  Damour, Henneaux and Nicolai
\cite{DamH1,DH1,DHN,DamH3,HPS} and some other authors. These works were
aimed from the very beginning at studying the generic behavior of
solutions near a spacelike singularity (a l\'a BKL) for
gravitational and cosmological models with non-diagonal metrics,
fields of forms and scalar fields. This approach was based on a wide use of Iwasawa decomposition
and hyperbolic Kac-Moody algebras \cite{Kac,Sac,BS,CCCMNNP}.
It was shown  in \cite{DamH3} that  for certain  models (of
supergravity) the billiards (or their parts)  are related to Weyl
chambers of certain hyperbolic Kac-Moody (KM) algebras.  This fact
has simplified the proof of the finiteness of the billiard volume
in certain cases.

 In the recent publications  \cite{KKN,KN,IMqb-1} the quantum billiard approach for the multidimensional
 gravitational model with several forms was considered.
 The main motivation for the quantum billiard approach in  \cite{KKN,KN} was coming from the quantum gravity paradigm;
 see \cite{N} and references therein.
  The asymptotic solutions to the WDW equation from \cite{KKN,KN}
 (in the model without scalar fields) are similar to those  obtained earlier in
 \cite{IMb0} for a multicomponent anisotropic fluid with
certain equations of state. In  \cite{IMqb-1}  another,
conformally covariant form of the WDW equation
\cite{Mis,Hal,IMZ,HK1,IMJ} was used. In this case the minisuperspace was enlarged
by including the form potentials for electric non-composite branes.
  In \cite{IMqb-1}  an example of a $9$-dimensional quantum  billiard
 for $D = 11$ model with $120$ four-forms  which mimic spacelike $M2$-brane solutions
 ($SM2$-branes in  $D=11$ supergravity) was considered. It was  shown that  the wave  function vanishes
 as $y^0 \to -  \infty$ (i.e. at the singularity),  where $y^0$ is the ``tortoise''
 timelike  coordinate in  minisuperspace \cite{IMqb-1}. In
 \cite{IMqb-2} we have generalized the approach of \cite{IMqb-1} to the case when
 scalar  fields with dilatonic couplings were added into consideration and
  the composite electromagnetic ansatz for branes was considered instead of
 the non-composite electric one from \cite{IMqb-1}.  New
 examples of  quantum billiards with electric and magnetic $S$-branes in $D=11$
 and $D=10$ models were presented. It was found that
in the quantum case  adding of magnetic  branes changes  the
asymptotic behavior of the wave functions, while
 it does not change the classical asymptotic oscillating behavior of the scale factors
 (and scalar field for $D=10$). It was found that in certain examples
  the basis wave  functions in the ``tortoise gauge'' vanish  as $y^0 \to -  \infty$.

 In this paper we generalize the approach from \cite{IMqb-2} to
 $n$  Einstein  factor spaces and a $\Lambda$ term. We also extend this approach
  by relaxing the main restriction for brane vectors: $(U^s, U^s) > 0$.
  Here we consider  examples of billiards in the model with $n$ non-intersecting
 electric branes, $n \geq 2$. The  brane world volumes are volume forms of $M_i$.
   We show that in the classical case any of these
 billiards describe the never ending oscillating behavior of the scale
 factors while approaching a singularity, which may be either
 spacelike or timelike one. The examples with timelike
 singularities are supported by the use of  either phantom form
 fields, or extra time-variables.
 For  $n=2$ (when 1-dimensional  $M_1$ and $M_2$ are forbidden) we obtain
 the basis asymptotic solutions to the WDW equation.

  We also generalize the model with $n$ electric branes  by adding a form  of rank $D$.
  This adding  does not change the  billiard but it  changes (e.g. drastically for $D \leq 7$) the
  basis asymptotic solutions to the WDW equation for a certain choice of Bessel function.

  Here we also consider an example of 4-dimensional quantum billiard  in $D =11$ model with ten 4-forms
   \cite{IMb1}.
  We use the ansatz with ten magnetic non-composite branes with brane worldvolumes of
  the form $M_i \times M_j \times M_k$   ($i < j < k$), where $M_i$ is $2d$ Einstein space, 
  $i = 1, \dots, 5$.
  We prove the vanishing of the basis wave  functions in the ``tortoise gauge''  as $y^0 \to -  \infty$.

  \section{The model}

 Here we study  the multidimensional gravitational
 model governed by the  action
 \bear{2.1}
  S_{act} = \frac{1}{2\kappa^{2}}
  \int_{M} d^{D}z \sqrt{|g|} {\cal L}    + S_{YGH},
 \ear
where
\beq{2.1L}
{\cal L} =
R[g]  - 2 \Lambda + h_{\alpha \beta}
  g^{MN} \partial_{M} \varphi^\alpha \partial_{N} \varphi^\beta
  \\ \nn
   - \sum_{a \in \Delta}
  \frac{\theta_a}{n_a!} \exp[ 2 \lambda_{a} (\varphi) ] (F^a)^2_g,
  \eeq
$g = g_{MN}(z) dz^{M} \otimes dz^{N}$ is a metric on the
manifold $M$, ${\dim M} = D$, $\Lambda$ is cosmological constant,
 $\varphi=(\varphi^\alpha)\in \R^l$
is a vector of scalar fields,
 $(h_{\alpha\beta})$ is a non-degenerate symmetric
 $l\times l$ matrix ($l\in \N$),
 $\theta_a  \neq 0$, and
 $$F^a =  dA^a   =\frac{1}{n_a!} F^a_{M_1 \ldots M_{n_a}}
  dz^{M_1} \wedge \ldots \wedge dz^{M_{n_a}}
 $$
 is an $n_a$-form ($n_a \geq 2$) on  $M$ and $\lambda_{a}$ is a
$1$-form on $\R^l$ :
 $\lambda_{a} (\varphi) =\lambda_{a \alpha}\varphi^\alpha$,
 $a \in \Delta$, $\alpha=1,\ldots,l$.
In (\ref{2.1})
we denote $|g| = |\det (g_{MN})|$,
  $(F^a)^2_g =
   F^a_{M_1 \ldots M_{n_a}} F^a_{N_1 \ldots N_{n_a}}
   g^{M_1 N_1} \ldots g^{M_{n_a} N_{n_a}},$
 $a \in \Delta$, where $\Delta$ is some finite set of (color) indices
 and $S_{\rm YGH}$ is the standard (York-Gibbons-Hawking) boundary term.
In the models with one time and the usual fields of forms all $\theta$ obey
$\theta_a > 0$ when the signature of the metric is $(-1,+1,
\ldots, +1)$. For such a choice of signature  $\theta_b < 0$
corresponds to a ``phantom'' form field $F^b$.

We consider the manifold
 \beq{2.10}
  M = \R_{*}  \times M_{1} \times \cdots \times M_{n},
 \eeq
with the metric
 \beq{2.11}
  g=  w e^{2{\gamma}(u)} du \otimes du   +
  \sum_{i=1}^{n} e^{2\beta^i(u)} g^i ,
 \eeq
where $\R_{*} = (u_{-}, u_{+})$, $w = \pm1$ and $g^i =
g^{i}_{m_{i} n_{i}}(y_i) dy_i^{m_{i}} \otimes dy_i^{n_{i}}$ is an
Einstein metric on $M_{i}$ satisfying the equation
 \beq{2.13}
  R_{m_{i}n_{i}}[g^i ] = \xi_{i} g^i_{m_{i}n_{i}},
 \eeq
 $m_{i},n_{i}=1, \ldots, d_{i}$; $\xi_{i}$ is constant,
 $i=1,\ldots,n$.
The functions $\gamma, \beta^{i} : \R_{*} \rightarrow \R $ are smooth.
We denote $d_{i} = {\rm dim} M_{i}$; $i = 1, \ldots, n$ and $d_0 = 1$;
 $D = \sum_{\nu = 0}^{n} d_{\nu}$.
We put any manifold $M_{i}$, $i = 1,\ldots, n$, to be oriented and connected.
Then the volume $d_i$-form
 \beq{2.14}
  \tau_i  \equiv \sqrt{|g^i(y_i)|}
   \ dy_i^{1} \wedge \ldots \wedge dy_i^{d_i},
 \eeq
and signature parameter
 \beq{2.15}
  \varepsilon(i)  \equiv {\rm sign}( \det (g^i_{m_i n_i})) = \pm 1
 \eeq
are correctly defined for all $i=1,\ldots,n$.

The cosmological ($S$-brane) solutions correspond to $w = - 1$ and positive
definite  $g^i$ for all $i$, while  static configurations (e.g. fluxbranes, wormholes, black branes etc.)
 may be obtained when $w = 1$, $g^k$ are Riemannian metrics for all $k > 1$ and
$g^1$ is the metric of pseudo-Euclidean signature $(-,+,...,+)$. Here we may also deal with solutions
having several timelike directions.

By $\Omega = \Omega(n)$  we denote a set of all non-empty subsets
of $\{ 1, \ldots,n \}$.
For any $I = \{ i_1, \ldots, i_k \} \in \Omega$, $i_1 < \ldots < i_k$,
we denote
 \bear{2.16}
  \tau(I) \equiv \tau_{i_1}  \wedge \ldots \wedge \tau_{i_k},
  \\
  \label{2.17}
  \eps(I) \equiv \eps(i_1) \ldots \eps(i_k),  \\
  \label{2.19}
   d(I) \equiv  \sum_{i \in I} d_i.
 \ear

For fields of forms we consider the following composite electromagnetic
ansatz:
 \beq{2.1.1}
  F^a= \sum_{I\in\Omega_{a,e}}{\cal F}^{(a,e,I)}+
          \sum_{J\in\Omega_{a,m}}{\cal F}^{(a,m,J)},
 \eeq
where
 \bear{2.1.2}
  {\cal F}^{(a,e,I)}=d\Phi^{(a,e,I)}\wedge\tau(I), \\
  \label{2.1.3}
  {\cal F}^{(a,m,J)}= e^{-2\lambda_a(\varphi)}*(d\Phi^{(a,m,J)}
  \wedge\tau(J))
 \ear
 are elementary forms of electric and magnetic types, respectively,
 $a\in\tri$, $I\in\Omega_{a,e}$, $J\in\Omega_{a,m}$ and
 $\Omega_{a,v} \subset \Omega$, $v = e,m$.
 In (\ref{2.1.3})
 $*=*[g]$ is the Hodge operator on $(M,g)$.

For scalar functions we put
 \beq{2.1.5}
   \varphi^\alpha=\varphi^\alpha(u), \quad
   \Phi^s=\Phi^s(u),
 \eeq
 $s\in S$. Thus, $\varphi^{\alpha}$ and $\Phi^s$ are functions on $(u_{-}, u_{+})$.

Here and below  the set $S$ consists of elements
\beq{2.1.5s}
s=(a_s,v_s,I_s),
\eeq
where $a_s \in \tri$ is the color index, $v_s = e, m$ is the electromagnetic
index, and the set $I_s \in \Omega_{a_s,v_s}$ describes the location
of the brane.

Due to (\ref{2.1.2}) and (\ref{2.1.3}) we get
  $d(I)=n_a-1, \quad d(J)=D-n_a-1$.

Here we present two restrictions on the sets of branes which
guarantee the diagonal form of the  energy-momentum tensor \cite{IMC}.

 The first restriction for a pair of two (different) branes both
 electric ($ee$-pair) or magnetic ($mm$-pair) with coinciding color index reads
   \beq{2.2.2a}
   d(I \cap J) \leq d(I)  - 2,
  \eeq
 for any $I,J \in\Omega_{a,v}$, $a\in \tri$, $v= e,m$ (here $d(I) =
 d(J)$).

  The second restriction for any pair of two branes
  with the same color index, which include one electric and one magnetic
  brane ($em$-pair) has the following form:
 \beq{2.2.3a}
   d(I \cap J) \neq 0,
  \eeq
 where  $I \in \Omega_{a,e}$, $J \in \Omega_{a,m}$, $a\in\tri$.

 These restrictions are satisfied identically
 in the non-composite case, when there are no two branes
 corresponding to the same form $F^a$ for any $a \in \tri$.

It follows from \cite{IMC} that the equations of motion for the model
 (\ref{2.1}) and the Bianchi identities,
   $ d{\cal F}^s=0$,
   $s \in S_m$, for fields from (\ref{2.11}),
  (\ref{2.1.1})--(\ref{2.1.5}), when restrictions ${\bf (R1)}$ and
${\bf (R2)}$ are  imposed, are equivalent to the equations of motion
for the $\sigma$-model governed by the action
  \beq{4.1.6}
   S_\sigma=\frac{\mu}{2} \int du{\cal N}\left\{
  {\cal G}_{\hat A\hat B}(X)\dot X^{\hat A}\dot X^{\hat B}
  - 2{\cal N}^{-2}V_w \right\},
  \eeq
where $X = (X^{\hat A})=(\beta^i,\varphi^\alpha,\Phi^s)\in
 {\R}^{N}$, $N = n +l + m$, $m = |S|$ is the number of branes
  and the minisupermetric    ${\cal G}=
   {\cal G}_{\hat A \hat B}(X)dX^{\hat A}\otimes dX^{\hat B}$
   on the minisuperspace  ${\cal M}= \R^{N}$ is defined as  follows:
 \beq{3.2.3n}
   ({\cal G}_{\hat A \hat B}(X))= (G_{ij}, h_{\alpha\beta},
   \eps_s \exp(-2U^s(\sigma))\delta_{ss'}).
 \eeq
where $\dot x\equiv dx/du$, $(\sigma^A)=(\beta^i,\varphi^\alpha)$,
$k_0 \neq 0$, the index set  $S$ is defined in (\ref{2.1.5s}),
 \beq{2.2.9}
  (\hat G_{AB})= {\rm diag}(G_{ij}, h_{\alpha\beta})
  \eeq
is the truncated target space metric with
  \beq{2.2.10}
   G_{ij}= d_i \delta_{ij} -d_i d_j,
  \eeq
and the co-vectors
 \bear{2.2.11}
  U^s =   U_A^s \sigma^A = \sum_{i \in I_s} d_i \beta^i -
  \chi_s \lambda_{a_s}(\varphi),
  \quad
  (U_A^s) =  (d_i \delta_{iI_s}, -\chi_s \lambda_{a_s \alpha}),
 \ear
 $s=(a_s,v_s,I_s)$,
 \beq{2.2.8}
  V_w =  - w \Lambda e^{2 {\gamma_0}(\beta)}
  + \frac{w}{2}   \sum_{i =1}^{n} \xi_i d_i e^{-2 \beta^i
  + 2 {\gamma_0}(\beta)}
 \eeq
is the potential with
 $\gamma_0(\beta)  \equiv \sum_{i=1}^nd_i\beta^i$,
and  ${\cal N}=\exp(\gamma_0-\gamma)>0$ is the modified lapse
function.

 We denote $\chi_e= +1$ and $\chi_m= -1$;
 \beq{2.2.12}
  \delta_{iI}=\sum_{j\in I}\delta_{ij}
 \eeq
 is the indicator of $i$ belonging
 to $I$: $\delta_{iI}=1$ for $i\in I$ and $\delta_{iI}=0$ otherwise; and
 \beq{2.2.13a}
   \eps_s=\eps(I_s) \theta_{a_s} \ {\rm for} \ v_s = e; \qquad
   \eps_s = -\eps[g] \eps(I_s) \theta_{a_s} \ {\rm for} \ v_s = m,
 \eeq
  $s\in S$, $\eps[g]\equiv\sign\det(g_{MN})$.

 In the electric case $({\cal F}^{(a,m,I)}=0)$ for finite internal space
volumes $V_i$ the action (\ref{4.1.6}) coincides with the action
(\ref{2.1}) if  $\mu=-w/\kappa_0^2$, $\kappa^{2} = \kappa^{2}_0 V_1 \ldots V_n$.

 In what follows we will use the scalar products
 of $U^s$-vectors $(U^s,U^{s'})$; $s,s' \in S$, where
 \beq{3.1.1}
  (U,U')=\hat G^{AB} U_A U'_B,
 \eeq
 for $U = (U_A), U' = (U'_A) \in \R^{N_0}$, $N_0 = n + l$ and
 \beq{3.1.2}
  (\hat G^{AB})= {\rm diag}(G^{ij}, h^{\alpha\beta})
   \eeq
is the matrix inverse to  the matrix
 (\ref{2.2.9}).
Here (as in \cite{IMZ})
 \beq{3.1.3}
    G^{ij}=\frac{\delta^{ij}}{d_i}+\frac1{2-D},
 \eeq
 $i,j=1,\dots,n$.

The scalar products (\ref{3.1.1})  read \cite{IMC}
 \beq{3.1.4}
  (U^s,U^{s'})=d(I_s\cap I_{s'})+ \frac{d(I_s)d(I_{s'})}{2-D}+
  \chi_s\chi_{s'}\lambda_{a_s \alpha} \lambda_{a_{s'}
  \beta} h^{\alpha \beta},
 \eeq
where $(h^{\alpha\beta})=(h_{\alpha\beta})^{-1}$ and
$s=(a_s,v_s,I_s)$,  $s'=(a_{s'},v_{s'},I_{s'})$ belong to $S$.

The potential (\ref{2.2.8}) reads  as follows:
 \beq{4.1.9}
  V_w=(-w\Lambda)\e^{2U^\Lambda(\sigma)}+\sum_{j=1}^n \frac{w}{2} \xi_j d_j
  \e^{2U^{(j)}(\sigma)},
 \eeq
where
 \bear{4.1.10}
   U^{(j)}(\sigma)=U_A^{(j)} \sigma^A=-\beta^j+\gamma_0(\beta),
   \qquad (U_A^{(j)}) =(-\delta_i^j+d_i,0),
   \\ \label{4.1.11}
   U^\Lambda(\sigma)=U_A^\Lambda \sigma^A=\gamma_0(\beta),
   \qquad (U_A^\Lambda)=(d_i,0).
 \ear

The scalar products of co-vectors $U^\Lambda$, $U^{(j)}$, $U^s$  are
defined by  the following relations \cite{IMC}
 \bear{4.1.14}
    (U^{(i)},U^{(j)})=\frac{\delta_{ij}}{d_j}-1, \qquad
    (U^{(i)},U^{\Lambda})=-1,  \qquad (U^{(i)},U^s)=-\delta_{iI_s},
    \\     \label{4.1.15}
    (U^s,U^{\Lambda})=\frac{d(I_s)}{2-D},
    \qquad  (U^{\Lambda},U^{\Lambda})=-\frac{D-1}{D-2},
 \ear
where $s=(a_s,v_s,I_s) \in S$;  $i,j= 1,\dots,n$.

The vector $U^\Lambda$ is a timelike
as well as $U^{(i)}$ with $d_i > 1$ (here we deal with $U^{(i)}$ obeying $\xi_i \neq 0$).
The vectors $U^\Lambda$ and $U^{(i)}$ with $d_i > 1$ belong to the same light cone (interiour part)
due to  relations $(U^{(i)},U^\Lambda)=-1$.

\section{Quantum billiard approach}

Here  we generalize the quantum  billiard
approach for  asymptotic solutions to the Wheeler-DeWitt (WDW) equation from
\cite{IMqb-2} to the case of a chain of Einstein spaces in the presence of the $\Lambda$-term.

 Let us denote by $S_{+}$ the subset of all $s \in S$ obeying
 \beq{3.1.uu}
   (U^s,U^{s})=    d(I_s) \left(1 +\frac{d(I_{s})}{2-D} \right) +
  \lambda_{a_s \alpha} \lambda_{a_{s}  \beta} h^{\alpha \beta} >0.
 \eeq

First we put the following additional restrictions on the model:
  \bear{3.1.4h}
  (i) \quad \quad \quad \quad \quad \quad  (h_{\alpha \beta}) > 0, \\ \label{3.1.4e}
  (ii) \quad \ \eps_s > 0 \quad {\rm  for \ all } \ s \in S_{+}.
    \ear

 These restrictions  are necessary conditions for the formation
 of infinite ``wall'' potential in hyperbolic spaces
 in certain limit (see below).
  The first restriction excludes  phantom scalar fields.
We note that in our previous work  \cite{IMqb-1,IMqb-2,IM-KPS-100} we used a more rigid restriction:
$S = S_{+}$.

 By  fixing the temporal gauge:
 \beq{4.2.1}
  \gamma_0-\gamma= 2f(X),  \quad  {\cal N} = e^{2f},
 \eeq
where $f$: ${\cal M}\to \R$ is a smooth function,  we obtain
the Lagrange system with the Lagrangian
 \beq{4.2.3}
   L_f = \frac{\mu}{2} \e^{2f}{\cal G}_{\hat A\hat B}(X)
   \dot X^{\hat A}\dot X^{\hat B} - \mu \e^{-2f}V_w
 \eeq
and the energy constraint
 \beq{4.2.4}
  E_f = \frac{\mu}{2} \e^{2f}{\cal G}_{\hat A\hat B}(X)
  \dot X^{\hat A}\dot X^{\hat B} + \mu\e^{-2f}V_w =0.
 \eeq

The set of Lagrange equations with the constraint (\ref{4.2.4})
is equivalent to the set of Hamiltonian equations for the
Hamiltonian
 \beq{4.2.3h}
   H^f=\frac{1}{2 \mu} \e^{-2f}{\cal G}^{\hat A\hat B}(X)
    P_{\hat A}  P_{\hat B} + \mu\e^{-2f}V_w
 \eeq
 with the constraint
 \beq{4.2.4h}
   H^f= 0,
 \eeq
where $ P_{\hat A} = \mu \e^{2f}{\cal G}_{\hat A\hat B}(X)  \dot
X^{\hat B}$ are momenta (for fixed gauge) and $({\cal G}^{\hat
A\hat B}) = ({\cal G}_{\hat A\hat B})^{-1}$.

Here we use the  prescriptions of covariant and conformally
covariant quantization of the hamiltonian constraint $H^f= 0$
which was suggested initially by Misner \cite{Mis} and considered
afterwards in \cite{Hal,IMZ,IMJ} and some other papers.

We obtain the Wheeler-DeWitt (WDW) equation,
 \beq{4.2.5}
   \hat{H}^f \Psi^f \equiv
    \left(-\frac{1}{2\mu}\Delta\left[e^{2f}{\cal G}\right]+
    \frac{a}{\mu} R\left[e^{2f}{\cal G}\right] + \mu\e^{-2f}V_w
     \right)\Psi^f=0,
  \eeq
where
 \beq{4.2.5a}
  a=a_N= \frac{(N-2)}{8(N-1)},
 \eeq
  $N = n+l + m$.

Here $\Psi^f = \Psi^f(X)$ is the wave function corresponding to
the $f$-gauge (\ref{4.2.1}) and satisfying the relation

 \beq{4.2.7}
  \Psi^f= e^{bf} \Psi^{f=0}, \quad b = b_N =(2-N)/2.
 \eeq

 In (\ref{4.2.5}) we denote by $\Delta[{\cal G}^f]$ and
 $R[{\cal G}^f]$  the Laplace-Beltrami operator and the scalar
 curvature corresponding to the metric
 \beq{4.2.7G}
 {\cal G}^f =   e^{2f} {\cal G},
 \eeq
 respectively.

The Wheeler-DeWitt (WDW) equation (\ref{4.2.5})) is conformally covariant.
  This follows from (\ref{4.2.7}) and the  relation:
 \beq{4.2.8}
 \hat{H}^f =   e^{-2f} e^{bf} \hat{H}^{f=0} e^{-bf},
 \eeq
where the coefficients $a_N$ and $b_N$ are  well-known in the
conformally covariant theory of  scalar field.

 Now we put $f = f(\sigma)$ and denote
   \beq{4.2.9U}
   \bar{U}= \sum_{s\in S} \bar{U}^{s}, \qquad  \bar{U}^{s} = U^{s}(\sigma) - f
   \eeq
    and
     \beq{4.2.9G}
     \bar{G}_{AB} =  e^{2f} \hat{G}_{AB}, \qquad \bar{G}^{AB} =  e^{-2f}
     \hat{G}^{AB}.
     \eeq

   Here we deal with a special class of asymptotic solutions to
   the WDW equation. Due to restrictions (\ref{3.1.4h}) and (\ref{3.1.4e})
 the  (minisuperspace) metrics  $\hat{G}$, $\cal{G}$ have  pseudo-Euclidean
   signatures $(-,+,\ldots ,+)$. We put $f = f_0$, where
       \beq{5.1}
         e^{2f_0} = - (\hat{G}_{AB}\sigma^A \sigma^B)^{-1},
       \eeq
   and we impose  $\hat{G}_{AB}\sigma^A \sigma^B < 0$. With this choice we deal with the
   so-called ``tortoise'' time gauge.

   Here  we  use a diagonalization of  $\sigma$-variables
       \beq{5.1.z}
      \sigma^{A} = S^{A}_{a}z^{a},
       \eeq
   $a = 0, ..., N_0-1$,  with $N_0 = n +l$, obeying $\hat{G}_{AB} \sigma^{A} \sigma^{B}
    = \eta_{ab} z^{a}z^{b}$, where $(\eta_{ab}) =     {\diag}(-1,+1,\ldots ,+1)$.

    We restrict  the WDW equation to the lower light cone
    $V_{-} = \{z = (z^{0}, \vec{z}) | z^{0} < 0, \eta_{ab} z^{a}z^{b} < 0 \}$
   and  we introduce the Misner-Chitr\'e-like coordinates
    \bear{5.2z}
     z^{0} = - e^{-y^{0}}\frac{1 + \vec{y}^{2}}{1 -\vec{y}^{2}}, \\
     \label{5.2zz}
     \vec{z} = - 2 e^{-y^{0}} \frac{\vec{y}}{1 - \vec{y}^{2}},
     \ear
    where $y^{0} < 0$ and $\vec{y}^{2} < 1$.

In these variables we have $f_0 = y^{0}$.  In the following we use

   \beq{5.4}
      \bar{G} = - dy^{0} \otimes d y^{0} + h_L,
   \eeq
   where
   \beq{5.5}
       h_L =  \frac{4 \delta_{rs} dy^{r} \otimes dy^{s}}{(1 - \vec{y}^{2})^2},
  \eeq
  (the summation over $r,s = 1,..., N_0 -1$ is assumed). The
  metric $h_L $ is defined on the unit ball
  $D^{N_0 -1} = \{ \vec{y} \in \R^{N_0 -1}| \vec{y}^{2} < 1
  \}$.  $D^{N_0 -1}$  with the metric $h_L$ is a realization of
  the   hyperbolic space $H^{N_0 -1}$.

 For the wave function  we consider the  ansatz from \cite{IMqb-2}:
 \beq{5.15}
        \Psi^{f_0} = e^{C(\sigma)} e^{iQ_s \Phi^s} \Psi_{0,L}(\sigma),
 \eeq
  where
   \beq{5.7}
        C(\sigma) = \frac{1}{2} \bar{U} =
       \frac{1}{2}(\sum_{s\in S} U^{s}_A \sigma^A - m f_0),
      \eeq
  where
  $Q_s \neq 0$  and $e^{iQ_s \Phi^s} = \exp(i \sum_{s \in S} Q_s \Phi^s)$.

Repeating all calculations from \cite{IMqb-2} we get
  \bear{5.16}
  \hat{H}^{f_0} \Psi^{f_0} = \mu^{-1} e^{C(\sigma)}
  e^{iQ_s \Phi^s}  \left(-\frac{1}{2} \tri[\bar{G}]+ \qquad \qquad  \right.\\ \nn
    \left.  \frac{1}{2} \sum_{s \in S} Q_s^2 e^{-2f_0 + 2U^s(\sigma)}
    +    \delta V  + \mu^2 e^{- 2f_0} V_w  \right) \Psi_{0,L}=0,
  \ear
where
     \beq{5.13}
      \delta V = A e^{-2f_0} - \frac{1}{8} (n+l-2)^2
      \eeq
and
       \beq{5.14}
      A  =   \frac{1}{8(N-1)} [ \sum_{s, s' \in S}
      (U^s,U^{s'}) - (N - 2) \sum_{s \in S} (U^s,U^{s}) ].
      \eeq

Here and in what follows   $U^s(\sigma) = U^s_A \sigma^A$.

Now  we proceed with the study the  asymptotic solutions to WDW
equation in the limit $y^0 \to - \infty$. Due to (\ref{5.15}) and
(\ref{5.16}) this equation reads
  \beq{5.17}
    \left(-\frac{1}{2} \tri[\bar{G}]+
    \frac{1}{2} \sum_{s \in S} Q_s^2 e^{-2f_0 + 2U^s(\sigma)}
    + \delta V + \mu^2 e^{- 2f_0} V_w \right) \Psi_{0,L}=0.
  \eeq

Here and in what follows we put $Q_s \neq 0$ for all $s \in S_{+}$.

It follows from the analysis of   \cite{IMb1} that for a certain choice of diagonalization
(\ref{5.1.z}),
  \bear{5.18}
    \frac{1}{2} \sum_{s \in S} Q_s^2 e^{-2f_0 + 2U^s(\sigma)}
    \to V_{\infty} \\ \label{5.19}
    e^{- 2f_0} \mu^2 V_w  \to 0,
  \ear
as $y^0 = f_0  \to - \infty$.
   Here  $V_{\infty}$ is the potential of infinite walls which
   are produced by branes with $(U^s,U^s) > 0$:
      \beq{5.18a}
   V_{\infty} = \sum_{s \in S_{+}} \theta_{\infty}( \vec{v}_s^2 -1 - (\vec{y} - \vec{v}_s)^2),
      \eeq
  where we denote $\theta_{\infty}(x) = + \infty $, for $x \geq
  0$ and $\theta_{\infty}(x) = 0$ for $x < 0$. The vectors
  $\vec{v}_s$, $s \in S_{+}$, which belong to $\R^{N_0 -1}$ ($N_0 = n
  +l$),   are defined by
      \beq{5.20}
      \vec{v}_s =  -  \vec{u}^s/u^{s}_{0},
      \eeq
  where the $N_0$-dimensional vectors  $u^s = (u^{s}_{0},\vec{u}^s) = (u^{s}_{a})$
  are obtained from brane $U^s$-vectors using the  matrix  $(S^{A}_{a})$
  from  (\ref{5.1.z})
  \beq{5.21}
    u^{s}_{a} = S^{A}_{a} U^s_A.
  \eeq
    By definition of $S_{+}$ we get
  \beq{5.21a}
  (U^s,U^s) = -(u_{s0})^2 + (\vec{u}_s)^2 > 0
  \eeq
  for all $s \in S_{+}$. In what follows
  we use a diagonalization (\ref{5.1.z})  obeying
  \beq{5.21b}
   u^{s}_{0} > 0, \qquad u^{(i)}_{0} > 0
  \eeq
  for all $s \in S$ and all $i$ such that $\xi_i \neq 0$ (and hence $d_i \neq 1$),
where   $u^{(i)}_{a} = S^{A}_{a} U^{(i)}_A$ are diagonalized
  curvature $U$-vectors, $i = 1, \dots, n$.
 The diagonalization (\ref{5.1.z}) from
 \cite{IMb1} obeys these conditions and implies (\ref{5.18})
   and (\ref{5.19}).

The inverse matrix   $(S_{A}^{a}) =
  (S^{A}_{a})^{-1}$ defines
   the map which is inverse to (\ref{5.1.z})
   \beq{5.21c}
   z^{a} = S_{A}^{a} \sigma^{A},
   \eeq
  $a = 0, ..., N_0 -1$.
     The inequalities (\ref{5.21a})
  imply  $|\vec{v}_s| > 1$ for all $s \in S_{+}$. The potential $V_{\infty}$
  corresponds to the billiard $B$ in the hyperbolic
  space $(D^{N_0 -1}, h_L)$. This  billiard is an open domain in
 $D^{N_0 -1}$ obeying the a set of inequalities:
     \beq{5.22}
       |\vec{y} - \vec{v}_s| < \sqrt{\vec{v}_s^2 -1} = r_s,
     \eeq
  $s \in S_{+}$. The boundary of the billiard  $\partial B$ is formed by  parts of
  hyper-spheres with  centers in $\vec{v}_s$ and radii $r_s$.

   The conditions (\ref{5.21b}) are  obeyed for the
   diagonalization (\ref{5.21c}) with
     \beq{5.21cc}
     z^{0} = e_A \sigma^{A},
     \eeq
   where $e =  (e_A)$ is a normed timelike vector $(e,e) = - 1$ obeying $(e,U^{\Lambda}) < 0$ and
   $(e,U^s) < 0$ for all $s \in S $. Hence $(e,U^{(i)}) < 0$    for all $i$ obeying $\xi_i \neq 0$.
   Our choice in   \cite{IMb1} was  $e =  U^{\Lambda}/\sqrt{|(U^{\Lambda},U^{\Lambda})|}$.

       When all factor spaces $M_i$ are Ricci-flat, i.e. all $\xi_i = 0$,
      brane part of conditions   (\ref{5.21b})  may be relaxed, while the curvature part of
      these conditions should be omitted.
     In this case we obtain a more general definition of the billiard walls (e.g. for $u_{s0} \leq 0$) described in \cite{IMb-09}.

   Thus, as in \cite{IMqb-2}, we are led to the asymptotic relation for the function
   $\Psi_{0,L}(y^0,\vec{y})$
   \beq{5.23}
    \left(-\frac{1}{2}\tri[\bar{G}]+ \delta V \right) \Psi_{0,L}=0
    \eeq
    with the zero boundary condition $\Psi_{0,L|\p B} = 0$
    imposed.

     Due to (\ref{5.4}) we get  $\tri[\bar{G}] = - (\p_0)^2 +  \tri[h_L]$, where $\tri[h_L] = \Delta_{L}$ is
     the Lapalace-Beltrami operator corresponding to the  metric $h_L$.

     By separating the variables,
      \beq{5.24}
    \Psi_{0,L}= \Psi_{0}(y^0) \Psi_{L}(\vec{y}),
    \eeq
    we obtain  the following asymptotic relation (for $y^{0} \rightarrow -
    \infty$)
     \beq{5.25}
     \left( \left(\frac{\partial}{\partial y^{0}}\right)^{2}  +
     2Ae^{-2y^{0}} +  E - \frac{1}{4} (N_0 -2)^2 \right)\Psi_{0} =
     0,
     \eeq
      where
       \beq{5.26}
       \Delta_{L}\Psi_{L} = - E \Psi_{L}, \qquad  \Psi_{L|\p B}=0.
       \eeq

       We assume that the minus Laplace-Beltrami operator $(-\Delta_{L})$  with the zero
       boundary conditions has a spectrum obeying  the following inequality:
       \beq{5.27}
        E \geq  \frac{1}{4} (N_0 -2)^2.
        \eeq
         This restriction was proved in  \cite{KKN,KN} for a wide class
        of billiards with finite volumes.

        Here we  restrict ourselves to the case of negative
        $A$-number        $A < 0$.

        Solving eq. (\ref{5.25}) we get for $A < 0$
        the following set of basis solutions:
       \beq{5.28}
       \Psi_{0} = {\cal B}_{i \omega} \left(\sqrt{2|A|}e^{-y^{0}}\right),
        \eeq
      where  ${\cal B}_{i \omega}(z) = I_{i \omega}(z), K_{i\omega}(z)$ are the modified Bessel
      functions and
      \beq{5.29}
       \omega = \sqrt{E -  \frac{1}{4} (N_0 -2)^2} \geq 0.
       \eeq

         We denote
            \beq{5.33}
            U(\sigma) = U_A \sigma^A =  \sum_{s \in S} U^{s}_A
            \sigma^A, \qquad U_A  =  \sum_{s \in S} U^{s}_A.
            \eeq

          In the following we  impose the restriction on $U = (U_A )$: $(U,U) < 0$.
         We have  $(U,U^{\Lambda}) < 0$  due to $(U^s,U^{\Lambda}) < 0$ for all $s$.

       From now we use a diagonalization with $z$-variables obeying    (\ref{5.21cc}) with
        \beq{5.e}
        e =  U/\sqrt{|(U,U)|}.
         \eeq
       For $U = k U^{\Lambda}$, with $k > 0$, such a diagonalization coincides with that of ref. \cite{IMb1}.

            It was obtained in \cite{IMqb-2} that
             \beq{5.35}
             \Psi^{f_0}
              \sim C_{\pm} \exp \left( \theta^{\pm}(|\vec{y}|)e^{-y^{0}} - \frac{1}{2}(m -1) y^{0}
              \right) e^{iQ_s \Phi^s} \Psi_{L}(\vec{y}),
              \eeq
             as $y^0 \to - \infty$  for any fixed $\vec{y} \in B$ and
          $C_{\pm}$ are non-zero    constants, ``plus'' corresponds to ${\cal B} = I$ and ``minus'' -
         to ${\cal B} = K$.
             Here
            \beq{5.36}
              \theta^{\pm}(|\vec{y}|) =
             - \frac{q}{2}   \frac{(1 + \vec{y}^{2})}{(1 -\vec{y}^{2})}
                                          \pm \sqrt{-2A},
              \eeq
              and
             \beq{5.34q}
             q = \sqrt{-(U,U)} > 0.
             \eeq

      Now we outline our analysis from \cite{IMqb-2} of asymptotic behavior of  $\Psi^{f_0}$
      as $y^0 \to - \infty$.  Here we fix all $\Phi^s \in \R$, $s \in  S$.

       For ${\cal B} = K$, $\Psi^{f_0} \to 0$
       as $y^0 \to - \infty$  for fixed $\vec{y} \in B$.

       Now let  ${\cal    B} = I$.

       If   $\frac{1}{2} q > \sqrt{2|A|}$,    or, equivalently,
        \beq{5.38U}
         \sum_{s \in S} (U^s,U^{s}) < -(U,U),
           \eeq
         we also get  $\Psi^{f_0} \to 0 $
        as $y^0 \to - \infty$  for fixed $\vec{y} \in B$.

        For $\frac{1}{2} q = \sqrt{2|A|}$,  or, equivalently,
         \beq{5.38Uu}
        \sum_{s \in S} (U^s,U^{s}) = -(U,U),
           \eeq
        we  get $\Psi^{f_0} \to 0 $
         as $y^0 \to - \infty$  for fixed $\vec{y}
       \in B \setminus \{ \vec{0} \} $. For $\vec{y} = \vec{0}$ we get $|\Psi^{f_0}| \to + \infty$
        as $y^0 \to - \infty$.
 It may be shown that in this case, when $m = N_0 = n + l$ (i.e. if $m$ is the minimal
number of walls which is necessary for the billiard to have a finite volume) we get
    \beq{5.38d}
  \Psi^{f_0}   \sim C_{0} \delta(\vec{y})  e^{iQ_s\Phi^s} \Psi_{L}(\vec{0}),
    \eeq
as  $y^0 \to - \infty$, where $C_{0} \neq 0$ is a constant
irrelevant for our consideration. Thus, for $m = n + l$ and for
eigenfunction   $\Psi_{L}(\vec{y})$ with  $\Psi_{L}(\vec{0}) \neq
0 $ we get a $\delta$-function in the asymptotic of  $\Psi^{f_0}$. In
this case we have an asymptotic localization of  $\Psi^{f_0}$ at the
point $\vec{y} = \vec{0}$ for our choice of gauge (``tortoise''
one). When the scalar fields are absent and we use a
diagonalization from \cite{IMZ,BIMZ} the relation $\vec{y} =
\vec{0}$ implies the isotropization $\beta^i = \beta$ and we may
talk in terms of asymptotic quantum isotropization of the wave
function in the temporal gauge under consideration.

     When $\frac{1}{2} q < \sqrt{2|A|}$,
             or, equivalently,
             \beq{5.38Ub}
        \sum_{s \in S} (U^s,U^{s}) > -(U,U),
      \eeq
       we get $|\Psi^{f_0}| \to + \infty$ as $y^0 \to - \infty$ for $\vec{y}$
     belonging to the open domain
     \beq{5.38B}
     B_{\infty} = \{ \vec{y} \in B:|\vec{y}| <
    \frac{2\sqrt{2|A|} - q}{2\sqrt{2|A|} + q} , \ \Psi_{L}(\vec{y}) \neq 0  \}.
     \eeq
     Outside the closure of $B_{\infty}$  we get the zero limit
     of our wave function and we  may
     talk in terms of the asymptotic localization
    of  $\Psi^{f_0}$ in $B_{\infty}$.

  With some exceptions we obtain the same
results for the asymptotic behavior  of the wave function in the
harmonic gauge with $f=0$:
$\Psi  =  e^{-by^0} \Psi^{f_0}$ in the limit $y^0 \to - \infty $,
 since the term $(-by^0)$ in the  exponent is suppressed generically by  $e^{-y^{0}}$. The change
 of gauge (from tortoise to harmonic)  may be sensitive for the asymptotic behavior  of the $\Psi$-function
 in the case (\ref{5.38Uu}) if $\vec{y} = 0$ and in the case (\ref{5.38Ub}) when
  $\vec{y}$ belongs to the border of the domain $B_{\infty}$.

\section{Example 1: $(n-1)$-dimensional  billiards in the models with electric branes}

Here we illustrate our approach by considering the model with the Lagrangian
 \bear{6.1}
  {\cal L} = R[g] - 2 \Lambda
   - \sum_{s =1}^{n}  \frac{\theta_s}{n_s!} (F^s)^2_g + \Delta  {\cal L}.
 \ear
Here we deal with the metric $g$  and the forms $F^s =  dA^s$, $s = 1, \dots, n$,
on the manifold $M$ from (\ref{2.10}). We use the metric ansatz from (\ref{2.11})
which deals with a warped product  of the interval $(u_{-}, u_{+})$ and $n$ Einstein spaces.
$\Delta  {\cal L}$ is an extra term with fields of forms which will be specified below.

 \subsection{The configuration with $n$ electric branes}

Here we put $\Delta  {\cal L} = 0$ and use the following non-composite electric ansatz for the fields of forms:
 \beq{6.1.2}
   F^{s}=d\Phi^{s}(u) \wedge \tau_s,
  \eeq
$s = 1, \dots, n$, where $n \geq 2$.

We put  $(U^s,U^{s}) > 0$ for all $s = 1, \dots, n$; by this we  exclude the
case $n=2$ with $(d_1,d_2) = (1,k), (k,1)$. The restriction (\ref{3.1.4e}) reads
\beq{6.1.4th}
    \theta_s \eps(s) > 0,
\eeq
$s = 1, \dots, n$.
According to these restrictions we get $\eps(s) = + 1$
for an ordinary form field  $F^s$ with $\theta_s > 0$
which means that the factor space $(M_s,g^s$) should be either Euclidean with the signature $(+,\dots,+)$, or
should have even number of timelike directions: $(-,-,+,\ldots,+)$ and so on.
For a phantom form field $F^s$ with $\theta_s < 0$ we should  consider the metric $g^s$ with
either pseudo-Euclidean  signature
$(-,+,\dots,+)$, or with the signatures $(-,\ldots, -,+,\ldots,+)$, containing  odd number of minuses.

There are three cases  here:  a) $w = -1$ in (\ref{2.11}) and all $(M_s,g^s)$ are Riemannian
spaces ($\theta_s  > 0$ for all $s$); b) $w = +1$,  $(M_1,g^1)$  has the signature $(-,-,+,\dots,+)$  and
 $(M_s,g^s$) are $s>1$ are Riemannian ($\theta_s  > 0$ for all $s$ );
c)  $w = +1$,  $(M_1,g^1$)  has the signature $(-,+,\dots,+)$  and  $(M_s,g^s$) with $s>1$ are Riemannian
($\theta_1  < 0$ and  $\theta_s  > 0$ for $s > 1$).
The case  a) describes cosmological solutions ($S$-branes), while  b) and c) may describe  static solutions,
e.g. with spherical, cylindrical,  and other symmetries.

For our configuration of branes (when $d_1 > 1$, $d_2 > 1$ for $n=2$)
 the billiard $B \subset H^n$ has a finite volume.  Indeed, let us suppose that $B$
has an infinite volume. Then there  exists a set of (real)
Kasner-like parameters   $\alpha = (\alpha_1$, ..., $\alpha_n)$
obeying the relations
 \beq{6.1.3}
    \sum_{i=1}^{n} d_i \alpha^i = \sum_{i=1}^{n} d_i (\alpha^i)^2  = 1,
  \eeq
and the inequalities  \cite{IMb1}
\beq{6.1.4b}
    U^s(\alpha) = d_s \alpha^s > 0,
  \eeq
 $s = 1, \dots, n$. Equations (\ref{6.1.3}) and (\ref{6.1.4b}) are not compatible:
otherwise  we get inequalities
 $0 < \alpha_s < (\alpha_s)^2  < 1$, for all $s$, which contradict  (\ref{6.1.3}).
This proves the finiteness  of the billiard  volume.

    Let us consider  the quasi-Cartan matrix  \cite{IMJ}
\beq{6.1.6}
  A_{s s'} = 2 (U^s,U^{s'}) / (U^{s'},U^{s'}),
  \eeq
where the scalar products (\ref{3.1.1})  read in our case
 \beq{6.1.4}
  (U^s,U^{s'})=d_s \delta_{s s'} - \frac{d_s d_{s'}}{D- 2},
  \eeq
  $s, s' = 1, \dots, n$.
Thus we are led to the matrix
\beq{6.1.6n}
 A_{s s} = 2, \qquad   A_{s s'} = - \frac{2 d_s}{D- 2 - d_{s'}}, \qquad s \neq s',
  \eeq
$s, s' = 1, \dots, n$.

It will be proved in a separate publication that the matrix (\ref{6.1.6n}) is coinciding with
the Cartan matrix of some hyperbolic Kac-Moody algebra in the following six cases
(up to permutations of indices): i) $n=2$, $(d_1,d_2) = (2,2),(2,3),(3,3)$,
ii) $n=3$, $(d_1,d_2,d_3)= (1,1,1), (1,1,2)$, iii) $n= 4$, $(d_1,d_2,d_3,d_4)= (1,1,1,1)$.

According to the   classification of hyperbolic KM algebras
by Carbone et al.  \cite{CCCMNNP}   only  the ranks $n=2,3,4$
should be considered here, since  for $n > 4$ there are no Dynkin diagrams where all nodes are connected by lines.

 For $n =2$ and  $(d_1,d_2) = (2,2), (2,3), (3,3)$ we get in (\ref{6.1.6n})
the Cartan matrices of the rank-2 hyperbolic KM algebras $H_2(p_1,p_2)$
with $(p_1,p_2) = (4,4), (4,3), (3,3)$, respectively.
Here we use the notation $H_2(p_1,p_2)$  for the hyperbolic KM algebra of rank 2
with the Cartan matrix  defined by the relations $A_{12} = - p_1$, $A_{21} = - p_2$,
where $p_1$ and $p_2$ are natural numbers obeying $p_1 p_2 > 4$.

For  $n =3$ and $(d_1,d_2,d_3) = (1,1,1)$ we obtain the hyperbolic
KM algebra by the number $7$ in the classification of  
Sa\c{c}lio\u{g}lu \cite{Sac} (see also  \cite{BS}), which is
number $80$ in the table  of ref.  \cite{CCCMNNP}.  In this case
 $A_{s s'} = - 2$ for all $s \neq s'$.  This KM algebra appears for Bianchi-IX
cosmology and its billiard  coincides with the Chitr\'e
one. In the quantum case this billiard was considered in numerous
papers; see \cite{GS,Kir-94,KKN,KN,Lec3} and  references therein.

For our model with a diagonal metric we may mimic the never ending
asymptotic behavior near the singularity for
 three scale factors  of Bianchi-IX model when
 $w = -1$,  $\eps(1) = \eps(2) = \eps(3) = +1$,   $\theta_1 = \theta_2 = \theta_3 = +1$.
In this case we deal with  approaching a spacelike
singularity in the $D=4$ model with
  three 2-forms. For  $w = +1$,  $\eps(1) = -1$, $ \eps(2) = \eps(3) = +1$,
$\theta_1 = -1$, $\theta_2 = \theta_3 =  +1$ we  find the never
ending asymptotic behavior of the scale factors near a timelike
singularity.  In this case we have a phantom 2-form $F^1$ and two ordinary $2$-forms $F^2$, $F^3$.

 For the case  $n=3$, $d_1 = d_2 = 1$, $d_3 =  2$ we get
 the billiard corresponding to the hyperbolic KM algebra by
number $40$ in the classification of  \cite{CCCMNNP}.
Here $A_{12} = A_{21} = - 1$, $A_{13} = A_{31} = A_{23} = A_{32} = - 2$.
We have a  billiard of finite volume which may describe the never ending
oscillating behavior near either spacelike or timelike
singularity. For the case of  a spacelike  singularity we put $w
= -1$ and use all metrics $g^i$  of Euclidean signatures and all
forms are taken to be ordinary ones. For the case with a timelike
singularity we have three (non-equivalent) possibilities with $w =
1$: (a)   $g^1 = - dx^1 \otimes dx^1$,  $g^2 = dx^2 \otimes dx^2$,
and $g^3$ has the signature $(+,+)$
 (b)   $g^1 = dx^1 \otimes dx^1$,  $g^2 = dx^2 \otimes dx^2$,
 and $g^3$ is of signature $(-,+)$;
(c)   $g^1 = dx^1 \otimes dx^1$,  $g^2 = dx^2 \otimes dx^2$, and
$g^3$ is of signature $(-,-)$. In the first two cases only one.
form should be phantom:  $F^1$ or $F^3$   in cases (a) or (b),
respectively. In the  case (c) all three forms are ordinary ones.

For the last example $n= 4$, $d_1 = d_2 = d_3 = d_4 = 1$ we get
the hyperbolic KM algebra by number 124 from  \cite{CCCMNNP}
with  $A_{ss'} = - 1$ for all $s \neq s'$. For
our model with four ordinary 2-forms  we get a diagonal cosmological
metric with $w= -1$ and $g^i = dx^i \otimes dx^i$, $i=1,2,3,4$,
which describes a never ending oscillating behavior near the
spacelike singularity. An analogous behavior will be obtained in
approaching a timelike singularity, if $w= 1$,  $g^1 = -
dx^1 \otimes dx^1$, $g^i = dx^i \otimes dx^i$, $i= 2,3,4$, when
the only one 2-form, namely $F^1$, is phantom.

For the model under consideration the basis asymptotic solutions
for the wave function  are given by eq. (\ref{5.15}) with the
 prefactor
  \beq{6.1.4C}
 C(\sigma) =   \frac{1}{2}(\sum_{s =1}^n d_s \beta^s - n y^0),
 \eeq
  and equations (\ref{5.15}), (\ref{5.24}), (\ref{5.26}), (\ref{5.28}), (\ref{5.29})
 where the relation for $A$-number (\ref{5.14}) reads
    \beq{6.1.4A}
    A  =   \frac{1}{8(2n-1)} \left[ - \frac{D-1}{D-2} -
    (2n - 2)  \sum_{s =1}^{n}  d_s \left(1 - \frac{d_s}{D -2} \right) \right] < 0.
       \eeq

Since our diagonalization  (\ref{6.1.4z}) uses a timelike
co-vector $U^{\Lambda}$ which coincides with sum of $n$ brane
vectors $U = U^1 + \cdots + U^n$, the whole of  our analysis of the asymptotic behavior
from the previous section is relevant.

We get
$|\Psi^{f_0}| \to 0$ as $y^0 \to - \infty$ for the basis
solutions with modified Bessel function ${\cal B} = K$.
For the basis solutions with another choice of modified Bessel function
  ${\cal    B} = I$ we  obtain  (generically) non-empty ``spots''  $B_{\infty} \subset B$
(see (\ref{5.38B})) for some basis  functions, where the  $|\Psi^{f_0}| \to + \infty$. These ``spots''
appear in the model under consideration since the inequality
\beq{5.1.4UU}
\sum_{s =1}^n (U^s,U^{s}) \geq -(U,U)
\eeq
 is valid   for  all sets $(d_1,d_2,\dots, d_n)$
with the exception:  $(d_1,d_2) = (1,k), (k,1)$. Indeed,  relation
(\ref{5.1.4UU}) in our case is equivalent to the relation
\beq{6.1.4d}
 (\sum_{s =1}^n d_s -2) \sum_{s =1}^n d_s  \geq
\sum_{s =1}^n d_s^2, \eeq
which could be readily proved for all
sets with the exception  $(d_1,d_2) = (1,k), (k,1)$. The equality
in (\ref{6.1.4d}) takes place only for $(d_1,d_2) = (2,2)$ and
$(d_1,d_2,d_3) = (1,1,1)$. In this case we may have a point-like
``spot'' for $\vec{y} = \vec{0}$ and a delta-function localization of
the wave function $\Psi^{f_0}$  for $y^0 \to - \infty$. This singularity can be
eliminated if we change to the harmonic gauge. In all other cases the
radius of any ``spot'' is non-zero and the ``spot'' cannot be eliminated by a transition
to the harmonic gauge. Equations (\ref{6.1.4psi}), (\ref{6.1.4E}), when
substituted into the general formulas of the previous section, will
give a solution to the problem in the quantum case.

Here the asymptotic solution to WDW equation are found up to the spectrum
of the (minus) Laplace-Beltrami operator (\ref{5.26})  with the zero
boundary conditions imposed. For the $n =2$ case this can be done explicitly.

{\bf The case $n =2$.} Now we consider the case $n=2$, when $d_1 \geq 2$, $d_2 \geq 2$.
We use  the following diagonalization of variables:
\beq{6.1.4z}
z^0 = q^{-1} (d_1 \beta^ 1 + d_2 \beta^2), \qquad  z^1 = q_1^{-1}
(\beta^ 1 -   \beta^2), \eeq
where $q = [(D - 1)/(D -2)]^{1/2}$ and
$q_1 = [(D - 1)/(d_1 d_2)]^{1/2}$. The components of $U^s$-vectors
in $z$-variables read \beq{6.1.4u} u^s_0  = \frac{d_s}{(D-2)q},
\qquad u^1_1 =  q_1^{-1}, \qquad u^2_1 = - q_1^{-1}. \eeq 
For 1-dimensional vectors from (\ref{5.20}) we get
 \beq{6.1.4v}
      v^1 =  -  R/d_1 < -1, \qquad  v^2 =    R/d_2 > 1.
      \eeq
 For $(d_1,d_2) = (2,2)$ we get $(v^1,v^2) = (-\sqrt{3}, \sqrt{3})$.
 Thus we are led to $1$-dimensional  billiard
$B = (y_1, y_2)$  with point-like walls assigned to
\beq{6.1.4y}
   y_1 = v^1 + \sqrt{(v^1)^2 - 1}, \qquad  y_2 = v^2 - \sqrt{(v^2)^2 - 1},
\eeq 
which obey   $-1 < y_1 < 0$ and $0 < y_2 < 1$. $B$ belongs
to the $1d$ unit ``disk'' $D^1 =(-1,1)$, which is an image of the
1-dimensional hyperbolic space $H^1 \subset \R^{1,1}$ under the
stereographic projection from the point $(z^0,z^1) = (-1,0)$. The
billiard is subcompact, i.e. its completion $[y_1, y_2]$ is
compact. We get $(y_1, y_2) = (-\sqrt{3} + \sqrt{2}, \sqrt{3} -
\sqrt{2})$  for $(d_1,d_2) = (2,2)$.

In the quantum case the model with two factor spaces is
integrable in the asymptotic regime of the formation of  billiard walls.
Here we have a discrete spectrum of  the Laplace-Beltrami
operator on $B = (y_1, y_2)$  with the metric $h_L =  4  dy
\otimes dy /(1 - y^{2})^{2}$ ($y = y^1$), when the zero boundary conditions at
points $y_1, y_2$ are imposed. Making the coordinate
transformation
   \beq{6.1.4x}
     x(y) = \ln \frac{1 + y}{1 -y},
    \eeq
we reduce the metric to the simple form $h_L =  dx \otimes dx $ and
$\Delta_{L} = d^2/dx^2$. We get a  discrete spectrum of the
Laplace-Beltrami operator on $(x_1 = x(y_1),x_2 = x(y_2))$  with the
zero boundary conditions:  $\Delta_{L} \Psi_{L,k} = -  E_k
\Psi_{L,k}$, $\Psi_{L,k}(x_i) = 0$, $i = 1,2$:
   \beq{6.1.4psi}
     \Psi_{L,k} = C_k \sin \left( k \frac{x - x_1}{x_2 - x_1} \pi \right),
     \quad  E_k = k^2 \pi^2/(x_2 - x_1)^2,
    \eeq
$k = 1,2,3, \dots$. Here constants $C_k \neq 0$ are irrelevant for
our consideration. The calculations give
\beq{6.1.4xv}
 x_2 - x_1 = \ln \frac{(1 + y^2)(1 - y^1)}{(1 - y^2)(1 + y^1)} = \frac{1}{2} \ln Q,
 \eeq
where
\beq{6.1.4Q}
  Q = \frac{( v^2 + 1)(|v^1| + 1 )}{(v^2 - 1)(|v^1| - 1)} =
  \frac{( R + d_1)(R + d_2 )}{( R - d_1)(R - d_2 ) }.
 \eeq

(For $(d_1,d_2) = (2,2)$ we get $x_2 - x_1 = \ln(2 + \sqrt{3})$.)
Hence the spectrum depends only on the parameter $Q$ and the quantum
number $k$:
 \beq{6.1.4E}
       E_k = \frac{4 \pi^2 k^2 }{(\ln Q)^2},
  \eeq
$k = 1,2,3, \dots$. In the symmetric case $d_1 = d_2 =d$
the eigenfunctions $\Psi_{L,k}$ have a zero at $x = 0$, or, equivalently,
at  $y = 0$,  only for even $k$. 

It may be shown that in the case $d_1 = d_2 =d \geq 4$ (i.e. for $D \geq 9$) the ``spot'' 
covers the billiard  with the exception of zeros of $\Psi_{L,k}$. This means
that $|\Psi^{f_0}(y)| \to + \infty$ as $y^0 \to - \infty$ for ${\cal B} = I$ and any $y$, obeying
$\Psi_{L,k}(y) \neq 0$. We recall that  
$|\Psi^{f_0}(y)| \to 0$ as $y^0 \to - \infty$ for ${\cal B} = K$ and any $y$. 
This is valid for $Q_1 \neq 0$,  $Q_2 \neq 0$.

In the classical case the model with two factor spaces is
also integrable in the asymptotic regime of the formation of  billiard walls.
This is considered in the Appendix in detail.

\subsection{The configuration with $n +1$ electric branes}

We extend the model from the previous subsection by adding an extra term which
  we  put in the Lagrangian (\ref{6.1}),
 \beq{6.2.1}
 \Delta  {\cal L} =  \frac{\theta_{0}}{n_0!} (F^0)^2_g,
 \eeq
 where  $F^0 =  dA^0$ is $D$-dimensional form, i.e. $n_0 = D$. We supplement
 the electric ansatz (\ref{6.1.2}) by the following relation
    \beq{6.2.2}
     F^{0}=d\Phi^{0}(u) \wedge \tau_1 \wedge ... \wedge \tau_n.
    \eeq

We get an additional brane vector   $U^0 = U^{\Lambda}$ and hence $U = U^0 + U^1 + \dots + U^n = 2 U^{\Lambda}$.
Since $(U^0, U^0) < 0$  adding the  term (\ref{6.2.1})
to the Lagrangian (\ref{6.1}) does not change the billiard and
the asymptotic behavior of the scale factors (near the singularity).

 Now the basis asymptotic solutions
 for the wave function  from the previous subsection are modified by adding the new variable $\Phi^{0}$
 and using another prefactor,
   \beq{6.2.4C}
  C(\sigma) =  \frac{1}{2}[2 \sum_{s =1}^n d_s \beta^s - (n + 1) y^0],
  \eeq
   and another $A$-number
     \beq{6.2.4A}
     A  =   \frac{1}{16n} \left[  (2n - 5) \frac{D-1}{D-2} -
     (2n -1)  \sum_{s =1}^{n}  d_s \left(1 - \frac{d_s}{D -2} \right) \right].
     \eeq
 In this case the relations $d_s \leq D -3$ (following from $(U^s,U^s) > 0$) imply
  $A \leq - \frac{1}{4n} \frac{D-1}{D-2} < 0$. Since the A-number from (\ref{6.2.4A})
  differs from (\ref{6.1.4A}) we are led to a different asymptotic
  behavior for the wave function $\Psi^{f_0} \to 0$ as $y^0 \to - \infty$ in this case
  when the electric brane with the brane vector $U^0$ obeying $(U^0, U^0) < 0$ is added.

 Now we consider  the relation for ``spots'' (\ref{5.1.4UU}).  We get
  \beq{6.2.4UU}
 \sum_{s =1}^n (U^s,U^{s}) + (U^{\Lambda},U^{\Lambda})  \geq - 4 (U^{\Lambda},U^{\Lambda}),
 \eeq
 or, equivalently,
  \beq{6.2.4D}
  (D-7)(D -1)   \geq  \sum_{s =1}^n d_s^2.
  \eeq

 In this case we get a  different restriction on ``spots'' in comparison with the relation
  (\ref{6.1.4d}). Indeed, due to (\ref{6.2.4D}) for $D \leq 7$  the ``spots''  are absent and
 hence   $|\Psi^{f_0}| \to 0$ as $y^0 \to - \infty$ for all basis solutions. Thus  the adding
 of the term (\ref{6.2.1}) to the Lagrangian changes (e.g. drastically) the asymptotic behaviour of
 the wave functions as $y^0 \to - \infty$, while in the classical case  this term is irrelevant
 for the asymptotic behaviour.

 \section{Example 2: $4$-dimensional billiard in $D=11$ model with ten magnetic branes}

Now we consider the $11$-dimensional model  with metric $g$ and ten $4$-forms
$F^J = d A^J$, $J \in \Omega$.
 The Lagrangian reads
\beq{6.3.1}
  {\cal L} =    R[g] - 2 \Lambda  -  \frac{1}{4!} \sum_{J  \in \Omega }  (F^J)^2_g.
   \eeq
   Here the index set $\Omega$ consists of all subsets  $J = \{j_1, j_2, j_3 \}  \subset \{1,2,3,4,5 \} $.
   The action (\ref{2.1}) is defined on the $11d$ manifold
   \beq{6.3.M}
     M = (u_{-}, u_{+}) \times M_{1} \times M_{2} \times M_{3} \times M_{4} \times M_{5},
    \eeq
    with $d_1 = d_2 =d_3 = d_4 =d_5 = 2$. We consider the cosmological
   ansatz for metric (\ref{2.11}) with $w = -1$ and five $2d$ Einstein spaces $(M_i,g^i)$
   of Euclidean signature $(+,+)$:
    \beq{6.3.g}
     g=  - e^{2{\gamma}(t)} dt \otimes dt   +
     \sum_{i=1}^{5} e^{2\beta^i(t)} g^i
    \eeq
   and impose the magnetic ansatz for the fields of forms
      \beq{6.3.2}
         F^{J}= *(d\Phi^{J} \wedge \tau(J)),
      \eeq
 $J \in \Omega$. Here  $\Phi^{J} = \Phi^{J}(t)$,  $*$ is the Hodge operator
 and $\tau(J) = \tau_{j_1}  \wedge  \tau_{j_2} \wedge \tau_{j_3} $,
   $J = \{ j_1, j_2, j_3 \} \in \Omega$, $j_1 < j_2 < j_3$. For $\Lambda = 0$ we deal with $10$
   non-composite magnetic branes which mimic $SM5$-branes in  ``truncated'' $D=11$ supergravity
   without Chern-Simons term.

   It was proved in \cite{IMb1} that the billiard $B \subset H^{4}$ has a finite volume.

   For all $J$ we get $d(J) = 6$, $(U^J,U^J) = 2$ and
   $U = \sum_{J } U^J = 6 U^{\Lambda} $ and hence $(U,U) = - 40$.
   Here $N =15$. We obtain
     \beq{6.3.4C}
    C(\sigma) =   6 \sum_{s =1}^5 \beta^s - 5 y^0,
    \eeq
    for prefactor  and
       \beq{6.3.4A}
       A  =   - \frac{75}{28} < 0
       \eeq
    for the $A$-number.

     In this case we get $\sum_{J } (U^J,U^J) < -(U,U)$ and
      hence    $|\Psi^{f_0}| \to 0$ as $y^0 \to - \infty$ for all basis solutions.
     In the harmonic time gauge we also get $|\Psi^{f = 0}| \to 0$ as $y^0 \to - \infty$ for all basis solutions.

    Thus in this example we have a similar asymptotic behavior of the basis wave functions to the case of
    $9d$ billiards  with a maximal number of $SM$-branes, either electric \cite{IMqb-1} or electric plus magnetic \cite{IMqb-2}.

\section{Conclusions}

 We have generalized the  quantum billiard approach from \cite{IMqb-2} by
considering a  cosmological-type model
with $n$ Einstein factor spaces in the theory with several
forms,  $l$ scalar fields and a $\Lambda$-term.
As in \cite{IMqb-2}, after imposing  the electromagnetic
composite brane ansatz with certain restrictions for brane
intersections and parameters of the model we have  used the
Wheeler-DeWitt (WDW) equation for the model, written in the
conformally covariant form.

By imposing   restrictions on the parameters of the model, e.g. on brane
$U^s$-vectors and
using the vanishing of the potential terms coming from curvatures
of the Einstein spaces and the $\Lambda$-term \cite{IMb1}, we have
 obtained  the asymptotic solutions to the WDW equation, in the
limit of  formation of  billiard walls: $y^0 \to - \infty$ ,
which have a form similar to that from \cite{IMqb-2}.

We have studied a subclass of examples of classical and
quantum billiards in the model with $n$ non-intersecting electric
branes, e.g., for certain hyperbolic KM algebras of ranks $n=2,3,4$.
In the classical case any of these billiards $B$ has a finite volume
and describes a never ending oscillatory
behavior of the scale factors while approaching  a singularity, which
may be either spacelike or timelike.

 In the quantum case the
asymptotic basis solutions to the WDW equation in the ``tortoise'' time gauge tends
to zero: $\Psi^{f_0} \to 0$, as $y^0 \to - \infty$, for the following choice of the Bessel function,
${\cal B} = K$, while  for another choice of the Bessel function: ${\cal B} = I$
we have obtained for some basis solutions $|\Psi^{f_0}| \to + \infty$ when $y^0 \to - \infty$
in certain domain $B_{\infty} \subset B$ of non-zero measure - ``spot'' -  for
all cases but $(d_1,d_2) = (2,2)$  and $(d_1,d_2,d_3) = (1,1,1)$. The
``spot'' does not disappear in the generic cases when the wave function in the harmonic time gauge is considered.
For  two  exceptional cases we have a point-like ``spot'' at $\vec{y} = \vec{0}$ for some basis solutions
 in ``tortoise'' time gauge, which  corresponds to  $\delta$-functions   in the limit $y^0 \to - \infty$,
  but this singularity does not take place in the harmonic time gauge.
 For  $n=2$ we have found the asymptotic solutions  for the metric in the classical case (see the Appendix) as
 well as its quantum counterpart, i.e. the asymptotic (basis) solutions to the WDW equation.

 Here we have  considered the branes with general scalar products $(U^s,U^s)$,
 while  in  \cite{IMqb-2} the restriction  $(U^s,U^s) > 0$
 was used.  The presence of branes
 with $(U^s,U^s) \leq 0$   has no effect on the billiard $B$ and the asymptotic
 classical behavior of  scale factors and scalar fields (as $y^0 \to - \infty$) but
 it changes the asymptotic solutions to WDW equation. We have illustrated
 this effect by an example with  $n$ electric  branes on product of $n$ Einstein  spaces
 obeying $(U^s,U^s) > 0$ and one brane obeying $(U^0,U^0) < 0$. It is shown that
 for such configuration of branes the ``spots'' disappear for $D \leq 7$
 and hence   $|\Psi^{f_0}| \to 0$ as $y^0 \to - \infty$ for all basis solutions to WDW
 equation. The analogous asymptotic behavior of all basis solutions  $\Psi^{f_0}$
 is shown to be valid for the $4$-dimensional quantum billiard in the $D=11$ model with ten magnetic branes,
 which was considered earlier for the classical case in \cite{IMb1}. This result can be extended
 to the configuration with composite magnetic S-branes in the model with one 4-form,
 but an open problem here is to include the Chern-Simons term
 of $D=11$ supergravity  into the consideration. In the classical case there were some obstacles for 
 doing so \cite{IMb1}.

 Recently, a certain interest in studying a
  possible oscillating behavior near a timelike singularity,
  started by  Parnovsky  \cite{Par1,Par2},   appeared after refs.  \cite{Kl,SW}.
    In  \cite{SW} the authors  speculated that such singularities,
  if occurring in AdS/CFT and being of the chaotic variety, may be
  interpreted as transient chaotic renormalization group flows
  which exhibit features reminiscent of chaotic duality cascades.
   So, the examples of billiards describing an oscillating behavior near a timelike
  singularity, which were considered in this paper,  may be tested   for a
  possible application  to the program suggested  in ref. \cite{SW}.

{\bf Acknowledgment}

This paper was financially supported by the Ministry of Education and Science
of the Russian Federation on the program to improve the competitiveness of
Peoples' Friendship University among the world's leading research and education centers in the
years 2016-2020.
The author thanks T. Damour for hospitality in
IHES (Bures-sur-Yvette), where this work was started, and also thanks H.
Nicolai for hospitality in AEI (Golm), where this work was
finished. The author thanks V.A. Belinskii and A. Kleinschmidt for useful
discussions.

\renewcommand{\theequation}{\Alph{section}.\arabic{equation}}
\renewcommand{\thesection}{}
\setcounter{section}{0}


\section{Appendix. Classical assymtotical solution for $n =2$.}


In classical case the motion of point-like particle in the billiard from
Section 4 describes asymptotic solution for the metric with ``jumping'' Kasner-like parameters
\beq{6.1.4g}
  g_{as} =  w  d \tau \otimes d \tau   +
  \sum_{i=1}^{n} B_i^2(\tau) \tau^{2 \alpha^i(\tau) } g^i, \quad w = \pm 1.
 \eeq
It is smooth when the synchronous-like variable $\tau$ belongs to
intervals $(\tau_1,\tau_2)$, $(\tau_2,\tau_3)$, $\dots$, where
$\tau_1 > \tau_2 > \tau_3 > \dots  > 0$ is an (in general unknown)
sequence of points tending to $0$, and the sets of real functions
$B(\tau) = (B_i(\tau) > 0)$, $\alpha(\tau) = (\alpha^i(\tau))$
take constant values on these intervals, i.e. $B(\tau) =
(B_i^{(k)})$, $\alpha(\tau) = (\alpha^i_{(k)})$ for  $\tau \in
(\tau_k, \tau_{k+1})$, $k = 1,2, \dots $, while the scale factors
$a_i(\tau) =  B_i (\tau) \tau^{\alpha^i(\tau) }$, $i=1,2$, are continuous
functions on $(\tau_1,0)$. All values $(\alpha^i_{(k)})$
obey the Kasner-like equations $(\ref{6.1.3})$.
The points $\tau_1,  \tau_2, \tau_3, \dots  > 0$
correspond to collisions with walls corresponding to branes. An impact
with  an $s$-wall in the billiard leads  to a  change of
the Kasner-like set $ \alpha \mapsto \hat{\alpha}$ \cite{Iv-02}.
 Here the following inequalities should be valid: $U^s(\alpha) = d_s \alpha^s < 0$ (before a collision with an   $s$-wall),
 $U^s(\hat{\alpha}) = d_s \hat{\alpha}^s > 0$ (after a collision with an  $s$-wall).

{\bf The case $n =2$.} Let us consider the case $n=2$, when $d_1 \geq 2$, $d_2 \geq 2$.
 We obtain  only two sets of solutions
to eqs. (\ref{6.1.3}) \cite{BIMZ}
\beq{6.1.4Kas}
  \alpha^1_{\pm} =   \frac{d_1  \pm \sqrt{R} }{d_1 (d_1 +   d_2)}, \qquad
  \alpha^2_{\pm} =   \frac{d_2  \mp \sqrt{R} }{d_2 (d_1 +   d_2)},
 \eeq
where
\beq{6.1.4R}
 R = \sqrt{d_1 d_2 (d_1 + d_2 -1)}.
 \eeq
Here the plus sign corresponds to the motion from the first wall with  $s= 1$  to the second one with
 $s= 2$, and the minus sign vice versa.

Let us consider the  billiard chamber $B_{ch}$, which is an open domain in the lower light cone
defined by the relations
   \beq{6.1.4uu}
   u^s_0 z^0 +  u^s_1 z^1 < 0, \ s= 1, 2,  \qquad z^0 < - |z^1|.
   \eeq
 By the transformations (\ref{5.2z}),  (\ref{5.2zz})  $B_{ch}$ is projected onto $B$. Any border line
  $L_s$ obeying $u^s_0 z^0 +  u^s_1 z^1 = 0$, or, equivalently
   \beq{6.1.4zc}
   z^1 =  c_s z^0, \qquad c_s = (v^s)^{-1},
   \eeq
is projected onto a point $y_s$, $s = 1,2$. The lines $L_1$ and
$L_2$ may be considered as world lines of point-like mirrors
moving with the velocities $ -1 < c_1 < 0$ and $ 0 < c_2 < 1$.
(Here, for the speed of light  we put $c = 1$.) The  asymptotic
motion  in the billiard $B$ with a sequence of bounce points:
$y_1$,$y_2$,$y_1$, $\dots$ is a projection of a zigzag world line
of a light beam in the billiard chamber  $B_{ch}$ with bounce
points: $z_1 \in L_1$, $z_2 \in L_2$, $z_3 \in L_1$, $\dots$.  For
the first part of our world line we get $z_2^0 - z_1^0 = z_2^1 -
z_1^1$   with $z^1_1 = c_1 z^0_1$ and $z^1_2 =  c_2 z^0_2$, which
implies
 \beq{6.1.4z1}
   z^0_2 =  Q_{+} z^0_1, \qquad Q_{+} = \frac{1 - c_1}{1 - c_2} > 1.
   \eeq
 For the second part of the world line we obtain
 $z_3^0 - z_2^0 =  -(z_3^1 - z_2^1)$   with  $z^1_3 =  c_1 z^0_3$
(and $z^1_2 =  c_2 z^0_2$) which gives us
  \beq{6.1.4z2}
   z^0_3 =  Q_{-} z^0_2 =  Q_{-} Q_{+} z^0_1, \qquad Q_{-} = \frac{1+ c_2}{1+ c_1} > 1.
   \eeq
Thus for $z^0$-components of bounce points we have
 \beq{6.1.4z0}
   z^0_1, \quad   z^0_2 =  Q_{+} z^0_1, \quad  z^0_3 =   Q_{-} Q_{+} z^0_1,
           \quad z^0_4 =  Q_{+} Q_{-} Q_{+}  z^0_1, \dots,
 \eeq
which may be written as follows
 \beq{6.1.4zs1}
   z^0_{2k +1} = Q^k z^0_1, \qquad  z^0_{2k} =  Q_{+} Q^{k-1} z^0_1, \qquad  Q = Q_{-} Q_{+} > 1,
   \eeq
for $k =1,2, \dots$.

In terms of dimensions $d_i$ parameters $Q_{-}, Q_{+}$ read
\beq{6.1.4Q}
   Q_{+} = \frac{R + d_1}{R - d_2} , \qquad Q_{-} = \frac{R + d_2}{R - d_1},
   \eeq
where $R = R(d_1,d_2)$ is defined in (\ref{6.1.4R}).

Now we use the  asymptotic relation for the volume scale factor $v =
\exp(d_1 \beta^ 1 + d_2 \beta^2)$ following from (\ref{6.1.4g}) and
(\ref{6.1.4z})
 \beq{6.1.4vq}
 v = \exp(qz_0) = C \tau,
 \eeq
where $C$ is the integration constant, $C= B_1^{d_1}
B_2^{d_2}$ for all  $\tau$. We put $\tau_1$ to be sufficiently
small, i.e. obeying at least  the restriction $C \tau_1 < 1$. Then
using  (\ref{6.1.4zs1}) and (\ref{6.1.4vq}) we get the asymptotic
relation for a set of bounce points   $\tau_1 > \tau_2 > \tau_3 >
\dots  > 0$, when $\tau_1$ corresponds to a collision with the
first wall:
\beq{6.1.4tau}
\ln(C \tau_{2k+1}) = Q^k \ln(C
\tau_1), \qquad \ln(C \tau_{2k}) = Q_{+} Q^{k-1} \ln(C
\tau_1),
\eeq
for $k =1,2, \dots$ and $C \tau_1 < 1$.

Using the  continuity conditions for the scale factors at the points $\tau_k$  we
get for $B(\tau_k -0) = B_i^{(k)}$, $i = 1,2$:
\beq{6.1.4B}
 B_i^{(2)} = \tau_2^{\Delta^i} B_i^{(1)}, \quad
B_i^{(3)} = \tau_3^{- \Delta^i} B_i^{(2)} = \tau_2^{ \Delta^i} \tau_3^{- \Delta^i} B_i^{(1)}, \dots
\eeq
where  $\Delta^i = \alpha^i_{+} - \alpha^i_{-}$, or
\beq{6.1.4D1}
 \Delta^1 = \frac{\Delta_0}{d_1}, \qquad
 \Delta^2 = - \frac{\Delta_0}{d_2}, \qquad  \Delta_0 = \frac{2 \sqrt{R}}{d_1 + d_2}.
\eeq

 We get from (\ref{6.1.4B})
\bear{6.1.4bo}
\ln(B_{i}^{(2k+1)}/ B_{i}^{(1)}) = \Delta^i Q_{+} (1 - Q_{-}) \frac{Q^k - 1}{Q - 1} \ln{(C \tau_1)},
\\ \label{6.1.4be}
\ln(B_{i}^{(2k)}/B_{i}^{(1)}) =
- \Delta^i \ln C  + \Delta^i [ (Q_{+} - 1 ) \frac{Q^k - 1}{Q - 1}  + 1 ] \ln{(C \tau_1)},
\ear
for  $i =1,2$ and $k =1,2, \dots$. Thus, we obtain an
asymptotic (oscillating) solution for the metric with two scale factors. It
 does not depend on $\Lambda$, the curvatures of the Einstein spaces and the brane charge densities. This approximation
works for small enough value of the parameter $C \tau_1 = \delta$.

 \end{document}